\documentstyle[12pt,aasms]{article}

\tighten

\begin{document}

\title{The Quasar Distribution in a Static Universe}
\author{David F Crawford}\
\affil{School of Physics A28,
University of Sydney, Sydney, N.S.W.  2006, Australia}

%\maketitle

\begin{abstract}
A crucial test of any cosmological model is the distribution of
distant objects such as quasars.
Because of well defined
selection criteria  quasars found by a ultraviolet excess (UVX)
survey are ideal candidates for testing the model out to a
redshift of $z = 2.2$.
The static cosmology proposed by
Crawford (1993) is used to analyse a recent quasar  survey
(\cite{BOYLE90}).
It is
shown that the distribution of number of quasars from the survey as
a function of redshift is in excellent agreement with the
predictions of the model.  A $V/V_m$ test on 351 confirmed quasars
with defined redshifts has a mean value of $0.568\pm 0.015$ with the
discrepancy being most likely due to incompleteness of the catalogue
at low redshifts.
For the redshift range $1.5 < z < 2.2$ where the accuracy of the cosmological
model is critical $V/V_m$ was $0.51 \pm 0.02$.
A well defined quasar luminosity function is derived that has a
peak at $M_B$ = -21.16 mag and is well fitted by a Gaussian
distribution in absolute magnitude with a standard deviation of
1.52 magnitudes.

\end{abstract}

\keywords{cosmology: quasars: quasar luminosity function}

\section{Introduction}
The  cosmological model  proposed by
Crawford (1993)
is stable and static with the Hubble redshift being due to a
gravitational interaction (\cite{CRAWFORD91}) that dissipates
photon energy into the inter-galactic plasma.  Since the model is
static the spatial density and any other characteristics of
objects must be independent of redshift; that is, there is no
possibility of evolution in any form.  As a result the model
makes simple and specific predictions about how the observed
distribution of objects should vary with redshift and magnitude.

Because of their high redshifts quasars are excellent objects for
probing the distant universe. Unfortunately most quasar surveys are
plagued with selection problems.  Most of these
have been overcome in a recent survey (\cite{BOYLE90}) in which the
COSMOS machine was used to find objects with an ultraviolet
excess (UVX) using UK Schmidt U and J plates.
They obtained spectra for 1400 objects using the Anglo-Australian
Telescope.  From their spectra 420 objects were identified as
quasars (including broad absorption line quasars).  Of these 351 that
had well defined redshifts with $z \leq 2.2$ were used as data
for this analysis.

In the next section, the relationship between apparent and
absolute magnitudes as a function of redshift is derived for
the static model.  Contrary to the standard (Big-Bang) model
(that apart from $H$) there are no free parameters, and thus
lack of agreement with the observations would seriously challenge
the static model. A
description of the observations is given, the
 $V/V_m$ statistic is determined and the data are
examined for any evidence of evolution.
A Monte Carlo calculation is used to show that even with optimum
maximum  likelihood
estimations using accessible volumes there is a remaining
bias in the estimated luminosity distribution.
Further analysis with allowance for this bias leads to a well
defined density distribution of absolute luminosity.
The general effects of selection bias are discussed in a simpler
context with only one region that has a single magnitude limit and
using these results it
is argued that the selection bias is important for all
cosmological models. The predicted number verses redshift
distribution of quasars is computed from the absolute luminosity
distribution and it is used to show that the estimation
of the luminosity distribution would improved by going to fainter
magnitudes.

\section{The cosmological model}
The cosmological model described in Crawford (1993) has the
same space geometry as the static Einstein universe (and incidentally the
same as a closed standard model that is not expanding) which is
that for a three dimensional surface of a four dimensional
hyper-sphere.
In this theory there is no universal expansion;
the Hubble redshift is caused by a gravitational
interaction with the inter-galactic plasma (\cite{CRAWFORD87A}).
The interaction cause the photon to lose energy via the emission of
very many very low energy secondary photons. In most cases the frequency
of the secondaries is well below the plasma frequency in which case
the energy loss is via direct transfer to plasma waves. In either case
the rate of energy loss is dependent
on the local plasma density.
The accumulated energy loss causes a redshift that is a function
 of the distance and the
density of the plasma.   Provided the plasma density does not
vary significantly the redshift is a good measure of the distance
to an object.  This is certainly true for large scales. For
smaller scales such as the size of galactic clusters there will
be variations in the plasma density that are reflected as
variations in the redshift. In this theory the voids and walls
seen in the distribution of galaxies are primarily due to
variations in the plasma density. However these variations are
small compared to the quasar redshift distances.

Since the cosmological model is static and is not evolving it
obeys the {\it perfect cosmological principle} (\cite{BONDI48}) in
which there is both spatial and time isotropy.  This does not
prevent objects such as galaxies or quasars from evolving.  What it
does require is that their creation and evolution is independent
of their location in space and  in time.  Hence any sampling of
the universe over a sufficiently large scale should not detect
any variation in the average characteristics of the objects such
as luminosity or density.
The application of this principle to quasars requires that their
spatial density and luminosity should be independent of their
redshift.  A critical test that would refute the static cosmology
is unequivocal evidence of evolution of luminosity or density of
of objects as a function of redshift.
Whereas the standard model not only permits evolution but since in it
galaxies have a finite lifetime and a common birth time,
 it demands evolution of galaxies
and presumably quasars. Lack of observed evolution in the standard model can
only arise by a fortuitous coincidence between the effects of  expansion
and galactic evolution.

If $R$ is the radius of the hyper-sphere that describes the
geometry of this static cosmology then the two-dimensional
surface area of the three-dimensional sphere of radius $r$ is
\begin{equation}
A(r) = 4\pi R^2 \sin^2(r/\!R),
\label{qso1}
\end{equation}
and the three-dimensional volume is
\begin{equation}
V(r) = 2\pi R^2 \left(r -
\frac{R}{2}\sin\left(\frac{2r}{R}\right)\right),
\label{qso2}
\end{equation}
where $r$ can vary from 0 to $\pi R$. Thus the total volume of the
universe on this model is $2\pi^2R^3$.

{}From Crawford (1991) the redshift ($z = \lambda_0/
\lambda_e - 1$) for a photon that has travelled a distance $r$ is
given by
\begin{equation}
 z = \exp\left(\frac{Hr}{c}\right) - 1,
\label{qso3}
\end{equation}
where $H$ is the Hubble constant. One of the major results of the
model is to relate the Hubble constant to the universe's radius
$R$. From Crawford (1993) we get $R = \sqrt{2}c/\!H$ which
provides the basic relationship
\begin{equation}
r = \frac{R}{\sqrt{2}}\ln(1 + z).
\label{qso4}
\end{equation}
Since the maximum value of $r$ is $\pi R$ equation~(\ref{qso4})
shows that the maximum value of z is 84.02 and that $z =
8.22$ when $r = \pi R/2$.

Suppose there are quasars (or any other objects) with spatial
density $\rho$ then the number in the interval $z$ to $z + dz$ is
given by
\begin{equation}
n(z) = \rho \frac{dV}{dz} = \rho \frac{dV}{dr} \frac{dr}{dz} =
\frac{2\sqrt{2}\pi R^3 \rho \sin^2(\ln(1 +z)/\!\sqrt{2})}{1 +z}.
\label{qso5}
\end{equation}
This distribution has a maximum value when $z = 2.861$.

Let a source of radiation have a luminosity $L(\nu)$
($\mbox{W.Hz}^{-1}$) at the emission frequency $\nu$.  Then if
energy is conserved the observed flux density $S(\nu_0)$
($\mbox{W.m}^{-2}\mbox{.Hz}^{-1}$) at a distance $r$  is the
 luminosity divided by
the area (equation~(\ref{qso1})) which is
\[
 S(\nu_0) = \frac{L(\nu)}{4\pi R^2 \sin^2(r/\!R)}.
\]
However because of the gravitational interaction there is an
energy loss such that the received frequency $\nu_0$ is related
to the emitted frequency $\nu_e$ by equation~(\ref{qso3}) and is
\[
   \nu_0 = \nu_e \exp(\sqrt{2}r/\!R) = \nu_e/(1 + z).
\]
The loss in energy means that the observed flux density is
decreased by a factor of $1 + z$.  But there is an additional
bandwidth factor $d\nu_e = (1 + z)\nu_0$ that tends to balance
the energy loss factor.  The balance is not perfect because the
source is observed at a different part of its spectrum from that
for a similar nearby source.  The correction for this spectral
offset is called the K-correction and Rowan-Robinson (1985) defines it
by
\[
 K(z) = -2.5\log\left(\frac{(1 + z)\int\psi(\nu_0)
L((1+z)\nu_0)d\nu_0}{\int\psi(\nu_0)L(\nu_0)d\nu_0}\right)
\]
where $\psi(\nu)$ is the filter transfer function and K is
expressed in magnitude units.  Note that the bandwidth correction is
explicitly included in the K-correction.  Then the
apparent magnitude $m$ is given by
\begin{eqnarray*}
m&  = &-2.5\log(S(\nu_0))\\
& = & -2.5\log(L(\nu_0)) + 5\log(\sin(r/\!R)) \\
 & & - 2.5\log(4 \pi R^2) + 2.5\log(1 + z) + K(z)
\end{eqnarray*}
Since the absolute magnitude $M$ is equal to the apparent magnitude at a
distance of $10 pc$ and using the relationship between $R$ and
$H$ we get
\[
M = -2.5\log(L(\nu_0)) -43.761 - 2.5\log(4 \pi R^2)
\]
where it is assumed that $H = 75\, \mbox{km.s}^{-1}\mbox{.Mpc}^{-1}$.

Hence the apparent magnitude is
\begin{equation}
m = M + 5\log(\sin(\ln(1+z)/\sqrt{2})) + 2.5\log(1 + z) + K(z) +
43.761.
\label{qso11}
\end{equation}
A common assumption  (\cite{BOYLE90}) is that on average the
continuum spectrum for quasars is of the form $L \propto \nu^{\alpha}$.
With this power law spectrum the K-correction is
\[
 K(z) = -2.5(1 + \alpha)\log(1 + z).
\]
Then the magnitude relation for a power law spectrum is
\begin{equation}
m = M + 5\log(\sin(\ln(1+z)/\sqrt{2})) - 2.5\alpha \log(1 + z)
+43.761.
\label{qso13}
\end{equation}

Let the density distribution of sources as a function of absolute
magnitude be $ f(M)$ such that the density of sources in
the range $M$ to $M + dM$ is
$d\rho = \rho_0  f(M)dM$
where $ f(M)$ is
normalized to one.  Then for $z$ in the range $z$ to $z + dz$ the
and using equation~(\ref{qso5})
the observed number of sources with apparent magnitudes in the
range $m$ to $m + dm$ is
\begin{equation}
 n(m)dm = 2\sqrt{2} \pi \rho_0 R^3 \int_{0}^{84.02} dz \frac{
  f(M(m,z)) \sin^2(\ln(1+z)/\sqrt{2})}{1 + z} dm
\label{qso14}
\end{equation}
where equation~(\ref{qso11}) or equation~(\ref{qso13}) is used to
calculate the absolute magnitude.  Thus the absolute luminosity
distribution and the K-corrections must be known before the
apparent magnitude distribution can be calculated.

\section{The observations}
As mentioned the set of observations used in this study is the
catalogue of quasars described by Boyle et al
(1990).  They used the COSMOS machine to measure the
apparent magnitudes on 8 sets of U and J plates taken with the UK
Schmidt telescope.  In this paper all the B magnitudes are their
uncorrected values and should be reduced by $0.1
\pm 0.05$ mag to get Johnston magnitudes.  The U magnitudes
were only used to get the \ub~ difference for object selection.
Objects were selected for further spectral analysis using the
FOCAP system on the Anglo-Australian Telescope.  This is a
multiple optical fibre spectrometer that could measure the spectra
of up to 64 objects in a 0.35 deg$^2$ field.  There were 34
fields spread over the 8 pairs of plates.  To be included an
object had to have an ultra-violet excess which meant that \ub~
was less than a value that was field dependent and which varied
from $-0.05$ to $-0.65$ mag.  Because of seeing and other
instrumental effects each field had its own magnitude limits,
typically from about 17 to 21 mag. After analysing the spectra they
found that 420 objects were quasars (including 9 broad emission
line objects) and most of the remainder were halo subdwarfs.  The
current analysis is restricted to a subset of 351 quasars each of
which had a well defined redshift.

Examination of the individual spectra provided by
Boyle et al (1990) shows that there is wide variation from a
constant spectral index  of $\alpha = -0.5$.  The major
requirement for the K-correction is in determining the B magnitude
at the emission wavelengths (near 4300\AA) from a value which has been
observed at
longer wavelengths.  Since the spectra covered the range from
3600\AA~ to 6600\AA~ they could be used to get an estimate of the
K-correction.  To do this it was assumed that the spectra could
be approximated by a power law and a set of templates were used
to estimate the power law index from the published spectra.  The
spectral indices estimated from the templates had a mean of
$-0.59$ and a standard deviation of $0.73$.  Although crude it is
slightly better than assuming a constant spectral index.

In order to calculate the accessible volume the
K-correction is also needed for wavelengths shorter than the B band.
In this case the \ub~ color index can be used to get a rough estimate
of the spectral index. By integrating the U and B response curves
(\cite{JOHNSON53}) multiplied by power laws it was found that a good estimate
of the power law index is given by $\alpha = -3.32 -4.0$
(\ub) where the constant term is chosen so that when
\ub~ = $-1.33$ mag the spectral index is 2.0.  In addition the
spectral indices estimated from the \ub~ color were restricted to
lie in the range $-2 \leq \alpha \leq 2$.  These spectral indices
had a mean of $-0.32$ and a standard deviation of 1.27 and for the
selected quasars the correlation coefficient between the two
indices was 0.7.  Thus the K-correction was estimated assuming a
power law spectrum with the index determined from template
matching to the published spectra for corrections to lower values
of $z$, and using the \ub~ spectral index for corrections to
higher values of $z$.

Boyle et al (1987) state that errors on the calibration of B
magnitudes range from $\pm 0.10$ mag at B = 18.0 mag to $\pm $
0.15 mag at B = 21.0 mag, with errors in \ub~ colors in the range
$\pm$ 0.15 mag to $\pm$ 0.25 mag. Of greater importance is the
completeness of the catalogue which is discussed extensively in
Boyle et al (1987). They claim a completeness of greater than 80\%
for quasars with $0.5 \leq z \leq 0.9$ and for the whole catalogue a
completeness of 90\%. Following their lead the analysis described
here was restricted to the range $0.3 \leq z \leq 2.2$
which reduced the number of quasars to 351.

\section{The $V/V_m$ test}
The $V/V_m$ test introduced by  Schmidt (1968) and
Rowan-Robinson (1968) is the average for all sources of
the ratio of the actual volume between the earth and the source
divided by the volume between the earth and the furthest point at
which the source would no longer satisfies the selection
criteria.  It is a statistic from the uniform distribution (0,1)
with a mean of 0.5 and a standard deviation for $n$ sources of
$1/\sqrt{12n}$.  For the data from Boyle et al (1990)
there are bright magnitude limits and the appropriate statistic
is $U/U_m$
(\cite{AVNI80}) which has the same characteristics as $V/V_m$.  If
the i'th quasar has a radial distance $r_i$ and its accessible
volume lies between the radii $a_i$ and $b_i$ then the $U/U_m$
statistic is defined by
\[
\left<\frac{U}{U_m}\right> = \frac{1}{n} \sum_{i=1}^{n}
\frac{v(r_i) - v(a_i)}{v(b_i) - v(a_i)}.
\]
where $v(r_i)$ is the volume
defined by
\begin{equation}
v(r_i) = \sum_{j} A_jV(r_{ij}),
\label{vol1}
\end{equation}
and where the index $j$ runs over those regions in which the quasar
could have been observed, $A_j$ is the solid angle of each region and
the volume $V(r_{ij})$ is computed using equation~(\ref{qso2}).
The $v(a_i)$ and $v(b_i)$ are computed using equation~(\ref{vol1})
except that the radii used are the minimum or maximum radii (respectively)
for which the quasar could have been observed in each region.
The absolute magnitude for each quasar was computed using
equation~(\ref{qso13}).
For the 351 selected quasars 198 had their
maximum volume determined by the z-limit.  That is they were bright
enough to be seen at $z=2.2$ and larger distances.

For 351 quasars the observed mean value is  $<U/U_m> = 0.569
\pm 0.015$ which is not in statistical agreement with the
expected value of 0.5. The observed standard deviation of
0.253 is in reasonable agreement with the expected value of
0.289.  However the number of quasars in each decile of $<U/U_m>$
which is given
in table~(\ref{tv1}) shows that the discrepancy is mainly
due to a deficiency of quasars in the first two deciles.
There appears to be a deficiency of about 50 weak nearby quasars, or
14\% of the total which is consistent with the limits of
completeness of the catalogue.  As a further check the data was
divided into two groups, 191 sources with $0.3 < z < 1.5$ and 160
sources with $1.5 < z \leq 2.2$.  For the first group $<U/U_m> = 0.58
\pm 0.02$
and for the second group it was $0.51 \pm 0.02$.  Since the differences
between cosmological models are negligible for low values of
redshift one would expect that if the model was incorrect the
higher redshift group (with $<U/U_m>$ = 0.51) would show the
largest deviation from 0.5.  The fact that it is the low redshift
group that shows the major discrepancy supports the argument that
the difference is mainly due to selection
effects.

In order to show that the $<U/U_m>$ ratios were not strongly dependent
on the K-corrections that were used the analysis was repeated with
a constant K-correction of $-0.5$.  For the range $0.3 < z < 2.2$
the result was $0.574$ to be compared with $0.569$ (above)R, and
for the same two groups the values were identical to those with a variable
K-correction.

Also shown in table~(\ref{tv1}) are the average redshift, average
apparent magnitude, the average absolute magnitude and the
average value of \ub~ for each decile.  As a function of $U/U_m$
there is a large (as expected) dependence of $z$ and a small, but
hardly significant (probability by chance of $\sim 5\%$),
variation in absolute magnitude and no apparent variation in
\ub~.
The expected result for a static cosmology is that any characteristic
of the quasars
should be independent of $U/U_m$.
Overall the results show that apart from small, possibly
selection, effects, the static cosmology is consistent with this
data on the $U/U_m$ test.

\section{The quasar luminosity function}
The accessible volume for each quasar ($V_i$) is defined using
equation~(\ref{vol1}) by
\[
V_i = v(b_i) - v(a_i).
\]
The luminosity distribution is obtained by selecting those
quasars within a small absolute magnitude range and dividing the
number of quasars by the average of their accessible volumes.  For
Poissonian statistics this is the maximum likelihood estimate.

Because of the strong selection effects on magnitude and redshift
the data only span part of the quasar luminosity range and as a
result the direct determination of the luminosity function is
biassed.  In order to determine this selection bias a Monte Carlo
program was used to simulate the selection effects.
The procedure was to choose a random absolute luminosity from a given
luminosity function and a random position within the maximum
accessible volume.
Next a spectral index from the uniform
distribution ($-2.0,2.0$) was chosen and
the \ub~ index was calculated
from $\ub~ = -0.25\alpha - 0.83$.
The apparent magnitude and redshift were then
calculated using equations~(\ref{qso3}) and (\ref{qso13}).
It was assumed that there is no correlation between absolute
magnitude and spectral index.  If this hypothetical object
satisfied the selection limits for one of the 34 areas it was
accepted for further processing.  Finally in order to simulate the
deficiency of weak sources the $V/V_m$ ratio was calculated and
if it was in the first decile 75\% of the objects were rejected
and if it was in the second decile 50\% were rejected.  The
simulated objects were then analysed by the same method used
to analyse the quasars.  The main purpose of this analysis was to
get density correction factors as a function of absolute
magnitude.  Although the accessible volume method corrects for
most of the selection effects the Monte Carlo analysis showed
that correction factors were still required to allow for the
a priori distribution of the quasar luminosities.
For example if the
parent luminosity distribution was a Gaussian in absolute
magnitude with a mean of $M_*= -21.15$ mag and with a standard
deviation in the range $1.0
\leq
\sigma_0 \leq 2.5$ mag then the observed
luminosity distribution is closely approximated by a Gaussian
with a standard deviation $\sigma = 0.27 + 0.73\sigma_0$ and with
an observed mean varying from $-22.33$ mag to $-22.92$ mag.  In effect the
selection effects truncate the distribution giving smaller
standard deviations.

Correction factors were calculated as a function of absolute
magnitude (in half magnitude steps) using a luminosity
distribution that is a Gaussian in absolute magnitude with
parameters ($-22.16$ mag, $1.52$ mag).  It should be noted that the
correction factors are only weakly dependent of the nature of
this parent distribution.
In fact the parent distribution is only required to determine
the correct weighting within each magnitude box.
However the factors are strongly dependent on the
magnitude and redshift selections.
The correction factors (that multiply the observed densities)
for $H = 75\, \mbox{km.s}^{-1}\mbox{.Mpc}^{-1}$ and for
three redshift ranges are shown in table~\ref{cf1}.

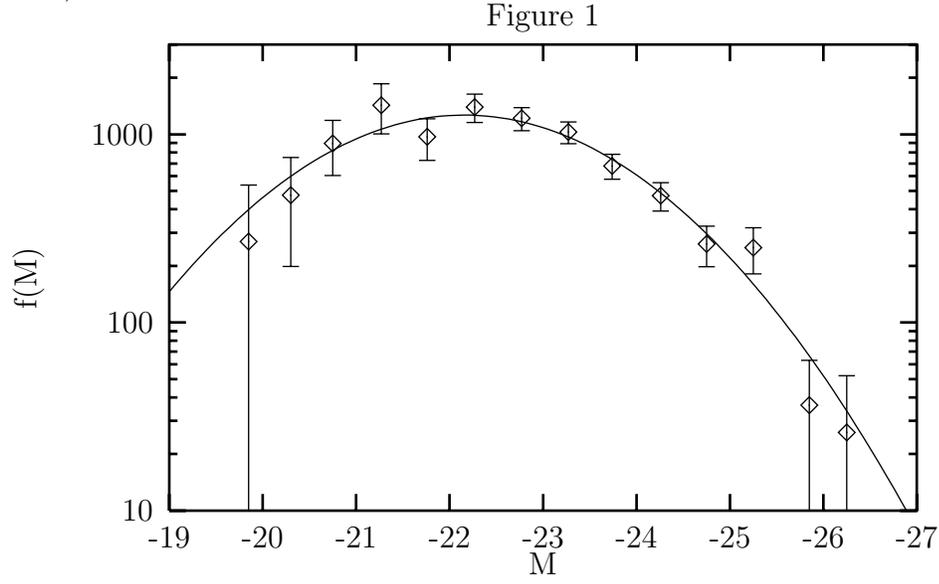
\begin{figure}[htb]
\caption{Luminosity function for UVX quasars as a function
\label{qlum.r}
of absolute magnitude. The unit is number per Gpc$^3$ per unit
magnitude interval. The curve is the best fit Gaussian (see text)}
%\plotone{qsof1.ps}
\epsfxsize=7cm

% GNUPLOT: LaTeX picture with Postscript
\setlength{\unitlength}{0.1bp}
\begin{picture}(3600,2160)(0,0)
\put(2008,2109){\makebox(0,0){Figure 1}}
\put(2008,51){\makebox(0,0){M}}
\put(100,1130){%
%
\makebox(0,0)[b]{\shortstack{f(M)}}%
%
}
\put(3417,151){\makebox(0,0){-27}}
\put(3065,151){\makebox(0,0){-26}}
\put(2713,151){\makebox(0,0){-25}}
\put(2361,151){\makebox(0,0){-24}}
\put(2009,151){\makebox(0,0){-23}}
\put(1656,151){\makebox(0,0){-22}}
\put(1304,151){\makebox(0,0){-21}}
\put(952,151){\makebox(0,0){-20}}
\put(600,151){\makebox(0,0){-19}}
\put(540,1670){\makebox(0,0)[r]{1000}}
\put(540,961){\makebox(0,0)[r]{100}}
\put(540,251){\makebox(0,0)[r]{10}}
\end{picture}
\end{figure}

The results for the density distribution (using the correction
factors) of the quasars as a function of absolute magnitude is shown
in figure~(\ref{qlum.r}).
The densities have been increased by a factor
of 416/351 (4 of the 420 quasars had $z >2.2$) to allow for
quasars that were not included.

Table~(\ref{tab0}) gives the magnitude range, the number of quasars,
the quasar density and the mean value of \ub~ for each of the
magnitude ranges.
The curve in figure~(\ref{qlum.r}) is the best fit Gaussian (chosen
for analytic simplicity) with the form
\begin{equation}
 f(M) = \frac{1}{\sqrt{2 \pi} \sigma}
\exp\left(-\left(\frac{M-M_*}{\sigma}\right)^2\right)
\label{lum1}
\end{equation}
where $M$ is the blue magnitude and since $ f(M)$ is normalized to
one it must be multiplied by the total quasar density $\rho_0$ to
get densities.  The values for the parameters for three values of
the Hubble constant are shown in table~(\ref{tab9}) where quoted
uncertainties are estimated standard deviations derived from the
Monte Carlos simulations.
These values were derived from independent
runs using different values for the Hubble constant and fully
include the effects of non-linearities in the equations.

What is most remarkable about this luminosity distribution is
that it has a well defined peak.
Whereas luminosity functions derived using the standard model
(\cite{HARTWICK90,BOYLE87}) rarely show a peak and are
spread over a much larger range in magnitudes.  Furthermore they
are different for different redshift ranges and there is little
agreement on the type of evolution that is needed to give
consistent results.  In contrast the static cosmology used here
has (apart from $H$) no free parameters and produces a well
defined luminosity function.  Since the observed luminosity
function will be broader than the actual luminosity function
because of uncertainties in the magnitudes, uncertainties
in the K-correction and because of
intrinsic variability in the magnitudes of quasars (\cite{HAWKINS93}),
the narrowness of the observed luminosity function suggests that
the contribution to its width from the errors in the cosmological model is
small and thus there is strong confirmation for the validity of
the static cosmological model.
As a further check the analysis was repeated with a constant K-correction
of $-0.5$ with the result that the  luminosity function was still peaked but
with a much larger width which shows that the use of the computed K-corrections
provides a worthwhile improvement.

In order to see if there are any evolutionary effects the
analysis for  the luminosity distribution  for three
redshift ranges with the same magnitude limits are shown
(for $H = 75\, \mbox{km.s}^{-1}\mbox{.Mpc}^{-1}$)
in table~(\ref{redz}) where the uncertainties ($\ 1 \sigma$)
have been estimated from repeated Monte Carlo simulations.
A statistical $t$ test shows that the differences in absolute
density between the near group and the far group is not
significant at the 5\% level and the differences in peak magnitude is
not significant at the 2.5\% level. Even if we admit a difference
in magnitude it could be easily due to the different effective magnitude ranges
or due to uncertainties in the
K-correction which has little effect on the nearby quasars and much
larger effects on the distant quasars.
Although there may be some effect due to deficiencies in the cosmological model
the reduction in the width of the peak for the two groups is mainly due
to the restricted magnitude range in each case.
The overall  result is that there is
no obvious evidence for evolution and that the static cosmological
model is consistent with the observations.

\section{Discussion}
\subsection{Selection bias}
The Monte Carlo results show that a strong selection bias still
exists even with maximum likelihood estimation of densities. The
cause of this bias is the non-linear relationship between
accessible volume, the absolute magnitude and the selected
apparent magnitude ranges. There are lesser dependencies on the
shape of the luminosity function and color selection (via the
K-correction).
Figure~(\ref{qwt}) shows the correction factor as a
function of magnitude for a simple example that has one region
with $0.0\leq z \leq 2.2$ and a single magnitude limit (with
$H = 75\, \mbox{km.s}^{-1}\mbox{.Mpc}^{-1}$ and variable
spectral index as described above).  Each curve is the Monte
Carlo results for a different magnitude cut-off.  Although these
results are computed for the static cosmological model it is
clear that the bias will exist for any cosmological model.
Repeated trials show that although
the complex behaviour seen on the left for the brighter limits
($B=19$ mag and $B=20$ mag) is genuine it should not be taken too
seriously since the
observed number of sources in this region is very low.
The
brightest peak in each curve corresponds to the absolute
magnitude of an object with the limiting apparent magnitude at a
redshift of about $z=0.36$.  Because of small numbers the bias to
the left of the brightest peak is very poorly determined.
Consider the curve with a cut-off at B$=20$ mag.  It has a peak
correction of 3.5 at $M_B = -20.8$ mag.  However in a survey only
about 3\% of the objects (within the full redshift range)
fainter than this would be detected.  It is because of the high
rejection rate that the non-linear selection effects become
important.  Again it should be emphasized that although the details
of the bias are dependent on the cosmological model the analysis
of any survey with any cosmological model should allow for
selection bias.

\subsection{The redshift distribution}
\begin{figure}[htb]
\caption{Number of quasars per square degree as a function of
\label{qzd}
redshift for magnitude limits of B=19,20,21,22,23,24 and 25.}
%\plotone{qsof3.ps}
\epsfxsize=7cm

% GNUPLOT: LaTeX picture with Postscript
\setlength{\unitlength}{0.1bp}
\begin{picture}(3600,2160)(0,0)
\put(2290,1816){\makebox(0,0)[l]{25}}
\put(2290,1713){\makebox(0,0)[l]{24}}
\put(2290,1482){\makebox(0,0)[l]{23}}
\put(2290,1118){\makebox(0,0)[l]{22}}
\put(2290,726){\makebox(0,0)[l]{21}}
\put(2290,443){\makebox(0,0)[l]{20}}
\put(2290,321){\makebox(0,0)[l]{19}}
\put(11924,304){\makebox(0,0)[l]{24}}
\put(11981,321){\makebox(0,0)[l]{23}}
\put(12037,336){\makebox(0,0)[l]{22}}
\put(11924,462){\makebox(0,0)[l]{21}}
\put(12488,409){\makebox(0,0)[l]{20}}
\put(13051,339){\makebox(0,0)[l]{19}}
\put(2008,2109){\makebox(0,0){Figure 3}}
\put(2008,51){\makebox(0,0){z}}
\put(100,1130){%
%
\makebox(0,0)[b]{\shortstack{n(z)}}%
%
}
\put(3417,151){\makebox(0,0){5.0}}
\put(3135,151){\makebox(0,0){4.5}}
\put(2854,151){\makebox(0,0){4.0}}
\put(2572,151){\makebox(0,0){3.5}}
\put(2290,151){\makebox(0,0){3.0}}
\put(2009,151){\makebox(0,0){2.5}}
\put(1727,151){\makebox(0,0){2.0}}
\put(1445,151){\makebox(0,0){1.5}}
\put(1163,151){\makebox(0,0){1.0}}
\put(882,151){\makebox(0,0){0.5}}
\put(600,151){\makebox(0,0){0}}
\put(540,2009){\makebox(0,0)[r]{30}}
\put(540,1716){\makebox(0,0)[r]{25}}
\put(540,1423){\makebox(0,0)[r]{20}}
\put(540,1130){\makebox(0,0)[r]{15}}
\put(540,837){\makebox(0,0)[r]{10}}
\put(540,544){\makebox(0,0)[r]{5}}
\put(540,251){\makebox(0,0)[r]{0}}
\end{picture}
\end{figure}
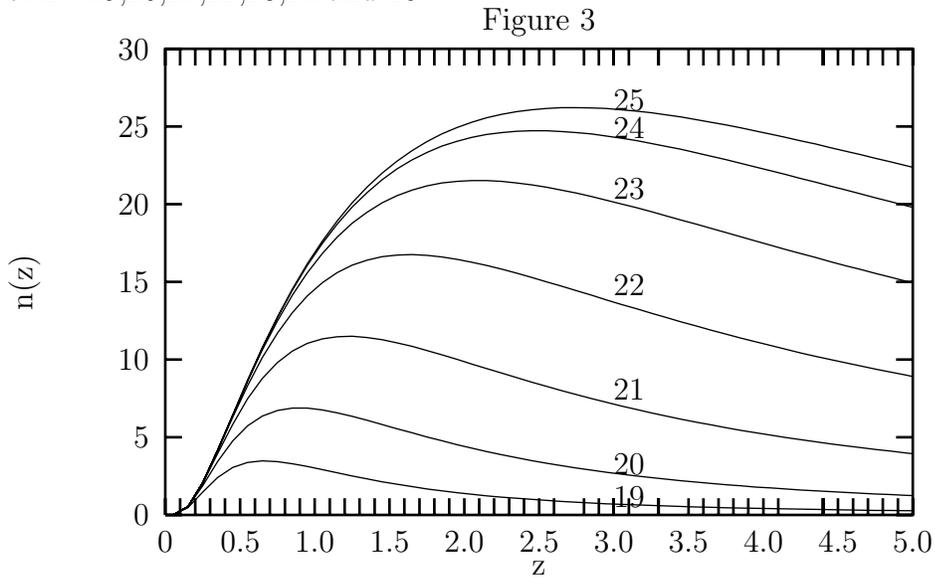

Given the luminosity distribution it a straight forward
integration using equation~(\ref{qso14}) to derive the number of
quasars that could be seen brighter that a magnitude limit as a
function of redshift.  Figure~(\ref{qzd}) shows the expected
number of quasars as a function of redshift for seven magnitude
limits.  The adopted luminosity distribution is a Gaussian in
absolute magnitude ($-21.16$ mag,$1.52$ mag) and the curves are calculated
for $H = 75\, \mbox{km.s}^{-1}\mbox{.Mpc}^{-1}$.
As the magnitude limit is
increased the curves peak at larger redshifts and asymptotically
approach equation~(\ref{qso5}).  Note that these are theoretical
curves and do not include the effects of selection bias and
therefore they cannot be directly compared with observational data.

\subsection{Further studies}
The success of the static cosmological model in producing a well
defined luminosity distribution enables one to use the model to
see what improvements could be made to future quasar surveys in
order to achieve a better luminosity function. Clearly the major
difficulty with current surveys is the limited magnitude range.
Figure~(\ref{qzd}) shows that at $z=2.0$ only about 36\% of the
quasars are observed with a magnitude less than $B=21$ and we
have to go to $B=24$ to get  94\% completeness. Thus it is clear that
determination of the luminosity distribution requires fainter
magnitude limits with well defined selection criteria. Whereas
tests of cosmological theories benefit more from larger redshift
ranges.
A second improvement would be better estimation of the
K-correction.  At large redshifts uncertainties in it easily
dominate the uncertainties in the absolute magnitude.
The ideal would be accurate integrations of each quasar spectrum
at zero redshift, the observed redshift and at the survey
limiting redshift.
Of further interest is that the Monte Carlo simulations show that
with larger numbers and a larger redshift range there is the
possibility of determining $H$ using quasar catalogues.
This is because the non-linearities in the magnitude-redshift
relationship and the volume-redshift relationship are becoming
important.

\section{Conclusion}
It has been shown that the model of a static universe is
consistent with the observed quasar distributions.
Subject to the uncertainty associated with the unknown
K-correction the $<U/U_m>= 0.568 \pm 0.015$, and the results
given in table~(\ref{tv1}) shows that apart from a lack of weak
nearby quasars there is no reason to reject the static cosmological
model with the $U/U_m$ test, in fact there is strong support.
The fact that the absolute magnitude density distribution shown
in figure(\ref{qlum.r}) has a well defined peak  is strong
evidence for the validity of the static model in explaining the
density distribution and magnitude relationship for quasars.
If the theory was seriously in error the peak would be very
broad or non existent.  Since this agreement of the static model
with the data is achieved without the fitting of any free
parameters it is strong support for the validity of the theory.
Furthermore it is apparent from this analysis that the major
deficiency in this data that limits its use as test for
cosmological theories is  its limited range in magnitudes and
uncertainties in the  K-corrections.

\section{Acknowledgements}
This work is supported by the Science Foundation for Physics
within the University of Sydney.

\clearpage

\begin{table}[htb]
\begin{center}
\caption{$U/U_M$ distributions
\label{tv1}}
\begin{tabular}{cccccc}
 decile & number & $<z>$& $<B>$ & $<M_B>$ & $<\ub~>$\\
\tableline
 ~1 & 10 &   0.556 & 18.74 &$-22.29$ &$ -0.74$  \\
 ~2 & 20 &   0.707 & 19.38 &$-22.55$ &$ -0.74$  \\
 ~3 & 35 &   0.950 & 19.62 &$-23.02$ &$ -0.83$  \\
 ~4 & 37 &   1.128 & 19.80 &$-23.08$ &$ -0.86$  \\
 ~5 & 39 &   1.224 & 19.98 &$-22.95$ &$ -0.75$  \\
 ~6 & 41 &   1.406 & 19.84 &$-23.07$ &$ -0.68$  \\
 ~7 & 45 &   1.514 & 20.24 &$-22.87$ &$ -0.77$  \\
 ~8 & 49 &   1.671 & 20.27 &$-23.08$ &$ -0.80$  \\
 ~9 & 38 &   1.806 & 20.30 &$-23.02$ &$ -0.59$  \\
 10 & 37 &   1.988 & 20.29 &$-23.45$ &$ -0.70$  \\
\end{tabular}
\end{center}
\end{table}

\clearpage

\begin{table}[htb]
\begin{center}
\caption{Density correction factors
\label{cf1}}
\begin{tabular}{cccccc}
 $<M_B>$ & $0.3\leq z \leq 2.2$ & $0.3\leq z \leq 1.5$ & $1.5\leq z \leq 2.2$\\
\tableline
$-19.25$ & 1.479 &  0.928  &       \\
$-19.75$ & 1.218 &  0.885  & 2.689 \\
$-20.25$ & 1.075 &  0.803  & 1.809 \\
$-20.75$ & 0.879 &  0.738  & 1.372 \\
$-21.25$ & 0.788 &  0.692  & 1.090 \\
$-21.75$ & 0.706 &  0.663  & 0.912 \\
$-22.25$ & 0.656 &  0.643  & 0.786 \\
$-22.75$ & 0.612 &  0.616  & 0.660 \\
$-23.25$ & 0.564 &  0.577  & 0.560 \\
$-23.75$ & 0.526 &  0.513  & 0.488 \\
$-24.25$ & 0.480 &  0.532  & 0.441 \\
$-24.75$ & 0.515 &  0.564  & 0.481 \\
$-25.25$ & 0.567 &  0.572  & 0.531 \\
$-25.75$ & 0.608 &  0.591  & 0.571 \\
$-26.25$ & 0.624 &  0.586  & 0.617 \\
\end{tabular}
\end{center}
\end{table}

\clearpage

\begin{table}[htb]
\begin{center}
\caption{Quaser luminosity function
\label{tab0}}
\begin{tabular}{llll}
  $<M_B>$ & count & $\rho_0 f /(\mbox{Gpc}^{-3}\mbox{.mag}^{-1})$ &
 $<\ub>$\\
$-19.85$  &    \hspace{1ex}1 &    \hspace{1ex}268$\pm   268$ & $-0.30$ \\
$-20.30$  &    \hspace{1ex}5 &    \hspace{1ex}475$\pm   277$ & $-0.72$ \\
$-20.75$  &    15 &    \hspace{1ex}895$\pm   291$ & $-0.73$ \\
$-21.27$  &    23 &    1429$\pm   324$ & $-0.69$ \\
$-21.76$  &    23 &    \hspace{1ex}968$\pm   241$ & $-0.61$ \\
$-22.27$  &    45 &    1396$\pm   141$ & $-0.80$ \\
$-22.77$  &    58 &    1216$\pm   170$ & $-0.77$ \\
$-23.27$  &    63 &    1027$\pm   134$ & $-0.79$ \\
$-23.74$  &    47 &    \hspace{1ex}679$\pm   102$ & $-0.75$ \\
$-24.26$  &    36 &    \hspace{1ex}471$\pm  \;\;80$ & $-0.81$ \\
$-24.75$  &    18 &    \hspace{1ex}261$\pm  \;\;63$ & $-0.74$ \\
$-25.25$  &    14 &    \hspace{1ex}249$\pm  \;\;68$ & $-0.59$ \\
$-25.85$  &    \hspace{1ex}2 &    \hspace{2ex}36$\pm  \;\;26$ & $-0.73$ \\
$-26.25$  &    \hspace{1ex}1 &    \hspace{2ex}26$\pm  \;\;26$ & $-0.50$ \\
\end{tabular}
\end{center}
\end{table}

\clearpage

\begin{table}[htb]
\begin{center}
\caption{Luminosity parameters for different Hubble constants
\label{tab9}}
\begin{tabular}{cccc}
$H/\mbox{km.s}^{-1}\mbox{.Mpc}^{-1}$  & $\rho _0/\mbox{Gpc}^{3}$ & $M_*$ &
$\sigma$\\
\tableline
  \hspace{1ex}50 & $\;1180 \pm \;58$ & $-22.94 \pm 0.12$ & $1.58 \pm 0.12$ \\
  \hspace{1ex}75 & $\;4820 \pm  240$ & $-22.16 \pm 0.12$ & $1.52 \pm 0.12$ \\
      100 & $ 13400 \pm  660$ & $-21.39 \pm 0.12$ & $1.62 \pm 0.12$ \\
\end{tabular}
\end{center}
\end{table}

\clearpage

\begin{table}[htb]
\begin{center}
\caption{Luminosity parameters for different redshift selections
\label{redz}}
\begin{tabular}{cccc}
selection  & $\rho _0/\mbox{Gpc}^{3}$ & $M_*$ & $\sigma$\\
\tableline
 $0.0 \leq z \leq 2.2$  & $4820 \pm 240$ & $-22.16 \pm 0.12$ & $1.52 \pm 0.12$
\\
 $0.0 \leq z \leq 1.5$  & $4570 \pm 510$ & $-22.06 \pm 0.22$ & $1.21 \pm 0.23$
\\
 $1.5 \leq z \leq 2.2$  & $3020 \pm 530$ & $-23.10 \pm 0.33$ & $1.33 \pm 0.27$
\\
\end{tabular}
\end{center}
\end{table}

\clearpage

\begin{figure}[htb]
\caption{Selection bias factor for a single region for magnitude limits
\label{qwt}
of B=19,20,21,22,23 and 24.}
%\plotone{qsof2.ps}
\epsfxsize=7cm

% GNUPLOT: LaTeX picture with Postscript
\setlength{\unitlength}{0.1bp}
\begin{picture}(3600,2160)(0,0)
\put(640,612){\makebox(0,0)[l]{24}}
\put(680,747){\makebox(0,0)[l]{23}}
\put(721,860){\makebox(0,0)[l]{22}}
\put(640,1829){\makebox(0,0)[l]{21}}
\put(1043,1423){\makebox(0,0)[l]{20}}
\put(1445,882){\makebox(0,0)[l]{19}}
\put(2008,2109){\makebox(0,0){Figure 2}}
\put(2008,51){\makebox(0,0){M}}
\put(100,1130){%
%
\makebox(0,0)[b]{\shortstack{Gain}}%
%
}
\put(3417,151){\makebox(0,0){-27}}
\put(3015,151){\makebox(0,0){-26}}
\put(2612,151){\makebox(0,0){-25}}
\put(2210,151){\makebox(0,0){-24}}
\put(1807,151){\makebox(0,0){-23}}
\put(1405,151){\makebox(0,0){-22}}
\put(1002,151){\makebox(0,0){-21}}
\put(600,151){\makebox(0,0){-20}}
\put(540,2009){\makebox(0,0)[r]{4}}
\put(540,1784){\makebox(0,0)[r]{3.5}}
\put(540,1558){\makebox(0,0)[r]{3}}
\put(540,1333){\makebox(0,0)[r]{2.5}}
\put(540,1107){\makebox(0,0)[r]{2}}
\put(540,882){\makebox(0,0)[r]{1.5}}
\put(540,657){\makebox(0,0)[r]{1}}
\put(540,431){\makebox(0,0)[r]{0.5}}
\end{picture}
\end{figure}
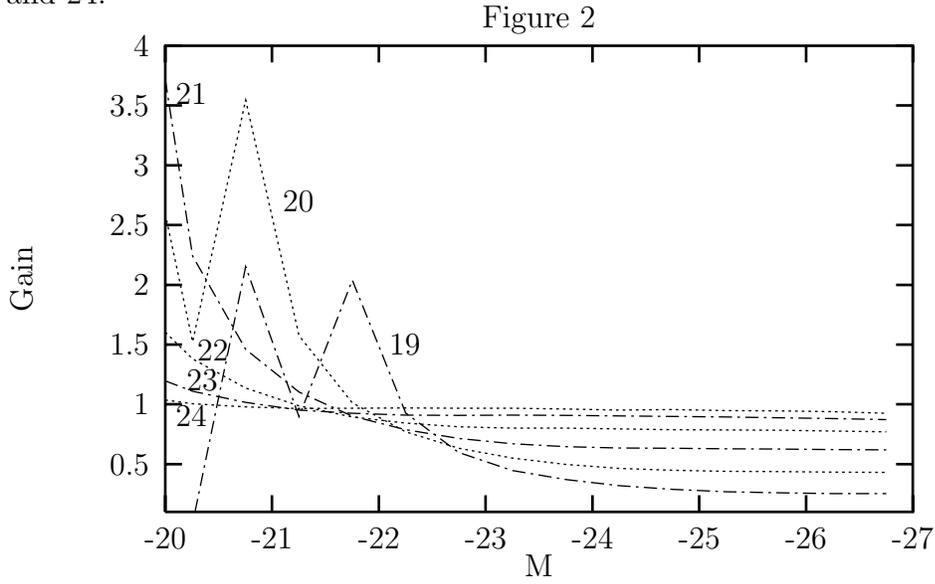

\end{document}